\documentclass[a4paper,twoside,10pt]{article}
\input{style.sty}
\begin{document}
%
\pagestyle{fancy}
\fancyhead{}
  \fancyhead[RO,LE]{\thepage}
  \fancyhead[LO]{T. Suzuki}          
  \fancyhead[RE]{Method of N-body simulation on the MOdified Gravity}
\rfoot{}
\cfoot{}
\lfoot{}
\label{P31}                             
\title{%
Method of N-body simulation on the MOdified Gravity
}
%
\author{%
  Takayuki Suzuki \footnote{Email address: n003wa@yamaguchi-u.ac.jp}$^{(a)}$
}
%
\address{%
  $^{(a)}$Department of Physics, Yamaguchi University, Yamaguchi city, 
       Yamaguchi prefecture 753-8512\\}
%
\abstract{
Scalar Tensor Vector Gravity(STVG) is one of modified gravity theories developed by John Moffat(2005).
MOG is abbreviated name for this theory.This theory is added a massive vector field to Brans-Dicke theory. 
It can explain a galactic rotation curve and the structure formation  without dark matter. It can also explain acceleration universe without dark energy.
However,these are claims by the developer and collaboraters.This theory was only inspected by simple approximate calculation.
Therefore it needs more objective verifications.We will carried out verification from the viewpoint of N-body simulation.
Such study is already accomplished by Brandao(2010).However, they did not precisely formulate N-body simulation on MOG.
This paper shows formulation of the N-body simulation on MOG more precisely.\\
}

\section{Introduction}
The standard model of cosmology today, the $\Lambda$CDM model, provides an excellent fit to cosmological observations.
 
But,most of the composition  of the universe {\em is invisible and undetectable which is named dark energy or dark matter}.
This fact provides a strong incentive to seek alternative explanations that can explain cosmological observations without dark matter or dark energy.

Modified Gravity (MOG) \citep{Moffat} has been used successfully to account for galaxy cluster masses \citep{Brownstein3}, the rotation curves of galaxies \citep{Brownstein}\citep{Brownstein2}, velocity dispersions of satellite galaxies, and globular clusters \citep{Toth3}. 
It can explain the observation of the Bullet Cluster \citep{Brownstein2} without cold dark matter.
Besides, MOG also meets the challenge posed by cosmological observations. 
In the paper(arXiv:0710.0364)\citep{MofTot2007},it is demonstrated that MOG produces an acoustic power spectrum, a galaxy matter power spectrum, and a luminosity-distance relationship that are in good agreement with observations.

However,these are claims by the developer and collaboraters.Therefore it needs objective verifications.
We want to carry out verification of MOG from the viewpoint of N-body simulation.
But,such study is already accomplished by Brandao et al.(2010)\citep{Araujo}.The result was negative for the Moffatian model.
They claim that the Moffatian potential cannot maintain exponential disks in dynamical equilibrium and therefore cannot be consistent with observations.
However, they did not precisely formulate N-body simulation on MOG.
This paper shows formulation of the N-body simulation on MOG in comformity with Moffat et al.(2009)\cite{Toth5}.

\section{Theory of MOG}
\subsection{Action}
The action of Moffat theory is constructed as follows \cite{Moffat}. 
STVG is formulated using the action principle. In the following discussion, a metric signature of $[+,-,-,-]$ will be used; the speed of light is set to $c=1$, and we are using the following definition for the Ricci tensor:
$R_{\mu\nu}=\partial_\alpha\Gamma^\alpha_{\mu\nu}-\partial_\nu\Gamma^\alpha_{\mu\alpha}+\Gamma^\alpha_{\mu\nu}\Gamma^\beta_{\alpha\beta}-\Gamma^\alpha_{\mu\beta}\Gamma^\beta_{\alpha\nu}.$
We denotes the Einstein-Hilbert Lagrangian:${\mathcal L}_G=-\frac{1}{16\pi G}\left(R+2\Lambda\right)\sqrt{-g},$
where $R$ is the trace of the Ricci tensor, $G$ is the gravitational constant, $g$ is the determinant of the metric tensor $g_{\mu\nu}$, while $\Lambda$ is the cosmological constant.
Introducing the Proca action Maxwell-Proca Lagrangian for the STVG vector field $\phi_\mu$:
\begin{eqnarray}
{\mathcal L}_\phi=-\frac{1}{4\pi}\omega\left[\frac{1}{4}B^{\mu\nu}B_{\mu\nu}-\frac{1}{2}\mu^2\phi_\mu\phi^\mu+V_\phi(\phi)\right]\sqrt{-g},
\end{eqnarray}
where $B_{\mu\nu}=\partial_\mu\phi_\nu-\partial_\nu\phi_\mu$, $\mu$ is the mass of the vector field, $\omega$ determines the strength of the coupling between the fifth force and matter, and $V_\phi$ is a self-interaction potential.
The three constants of the theory, $G$, $\mu$ and $\omega$, are promoted to scalar fields by introducing associated kinetic and potential terms in the Lagrangian density:
\begin{eqnarray}
{\mathcal L}_S=-\frac{1}{G}\left[\frac{1}{2}g^{\mu\nu}\left(\frac{\nabla_\mu G\nabla_\nu G}{G^2}+\frac{\nabla_\mu\mu\nabla_\nu\mu}{\mu^2}-\nabla_\mu\omega\nabla_\nu\omega\right)+\frac{V_G(G)}{G^2}+\frac{V_\mu(\mu)}{\mu^2}+V_\omega(\omega)\right]\sqrt{-g},
\end{eqnarray}
where $\nabla_\mu$ denotes covariant differentiation with respect to the metric $g_{\mu\nu}$, while $V_G$, $V_\mu$, and $V_\omega$ are the self-interaction potentials associated with the scalar fields.
The STVG action integral takes the form $S=\int{({\mathcal L}_G+{\mathcal L}_\phi+{\mathcal L}_S+{\mathcal L}_M)}~d^4x,$
where ${\mathcal L}_M$ is the ordinary matter Lagrangian density.

\subsection{Weak field approximation of MOG}
The field equations of STVG can be developed from the action integral using the variational principle.
First a test particle Lagrangian is postulated in the form${\mathcal L}_\mathrm{TP}=-m+\alpha\omega q_5\phi_\mu u^\mu,$
where $m$ is the test particle mass, $\alpha$ is a factor representing the nonlinearity of the theory, $q_5$ is the test particle's fifth-force charge, and $u^\mu=dx^\mu/ds$ is its four-velocity.
Assuming that the fifth-force charge is proportional to mass,$q_5=\kappa m$, the value of $\kappa=\sqrt{G_N/\omega}$ is determined 
and the following equation of motion is obtained in the spherically symmetric, static and weak gravitational field of a point mass of mass $M$:

\begin{eqnarray}
\ddot{r}=-\frac{G_NM}{r^2}\left[1+\alpha-\alpha(1+r/\lambda)e^{-r/\lambda}\right],\label{PointAcceleration}
\end{eqnarray}
where $G_N$ is Newton's constant of gravitation.
Further study of the field equations allows a determination of $\alpha$ and $\lambda$ for a point gravitational source of mass $M$ in the form:
$\lambda=1/\mu=\frac{\sqrt{M}}{D}$,$\alpha=\frac{19M}{(\sqrt{M}+E)^2}$
The constants $D$ and $E$ various astronomical observation yield the following values:
$D\simeq 6250 M_\odot^{1/2}\mathrm{kpc}^{-1},$ $E\simeq 25000 M_\odot^{1/2},$

In the weak-field approximation, STVG produces a Yukawa-like modification of the gravitational force due to a point source.
Nearby a source gravity,the repulsive force which comes from vector field counteracts attraction.
But far from a source gravity,Yukawa-like repulsive force cannot reach because vector field is massive.

Intuitively, this result can be described as follows: far from a source gravity is stronger than the Newtonian prediction, 
but at shorter distances,gravity is comparable to Newtonian.

\begin{figure}[h!]
\centering
\includegraphics[keepaspectratio=true,width=13cm]{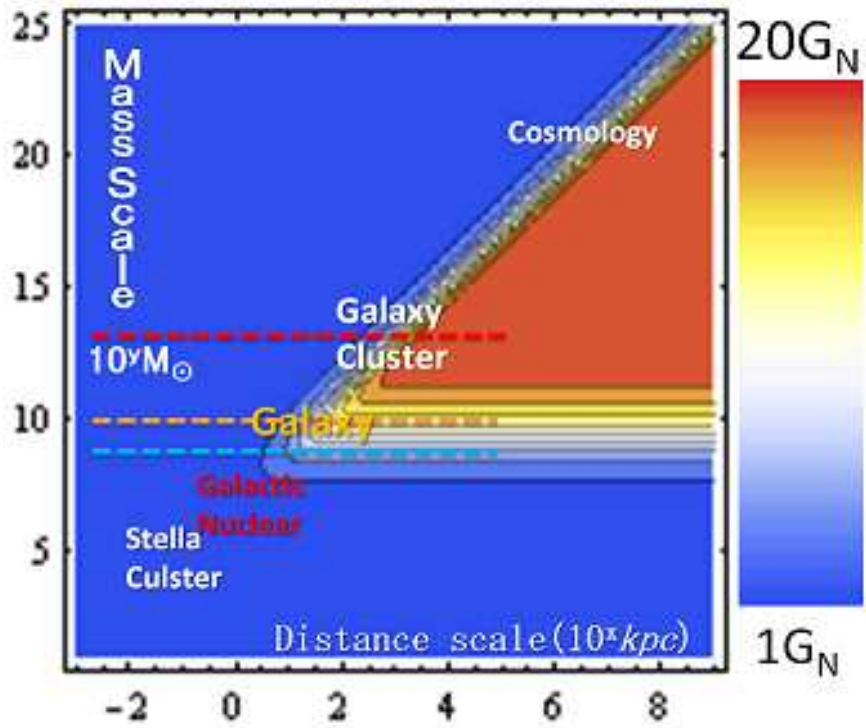}
\caption{Relation between scale and effective G. X-axis shows a distance scale.Y-axis shows mass of the gravity source. 
}
\label{fig:P31_ScaleVSG}
\end{figure}

\begin{figure}[h!]
\centering
\includegraphics[keepaspectratio=true,width=12.5cm]{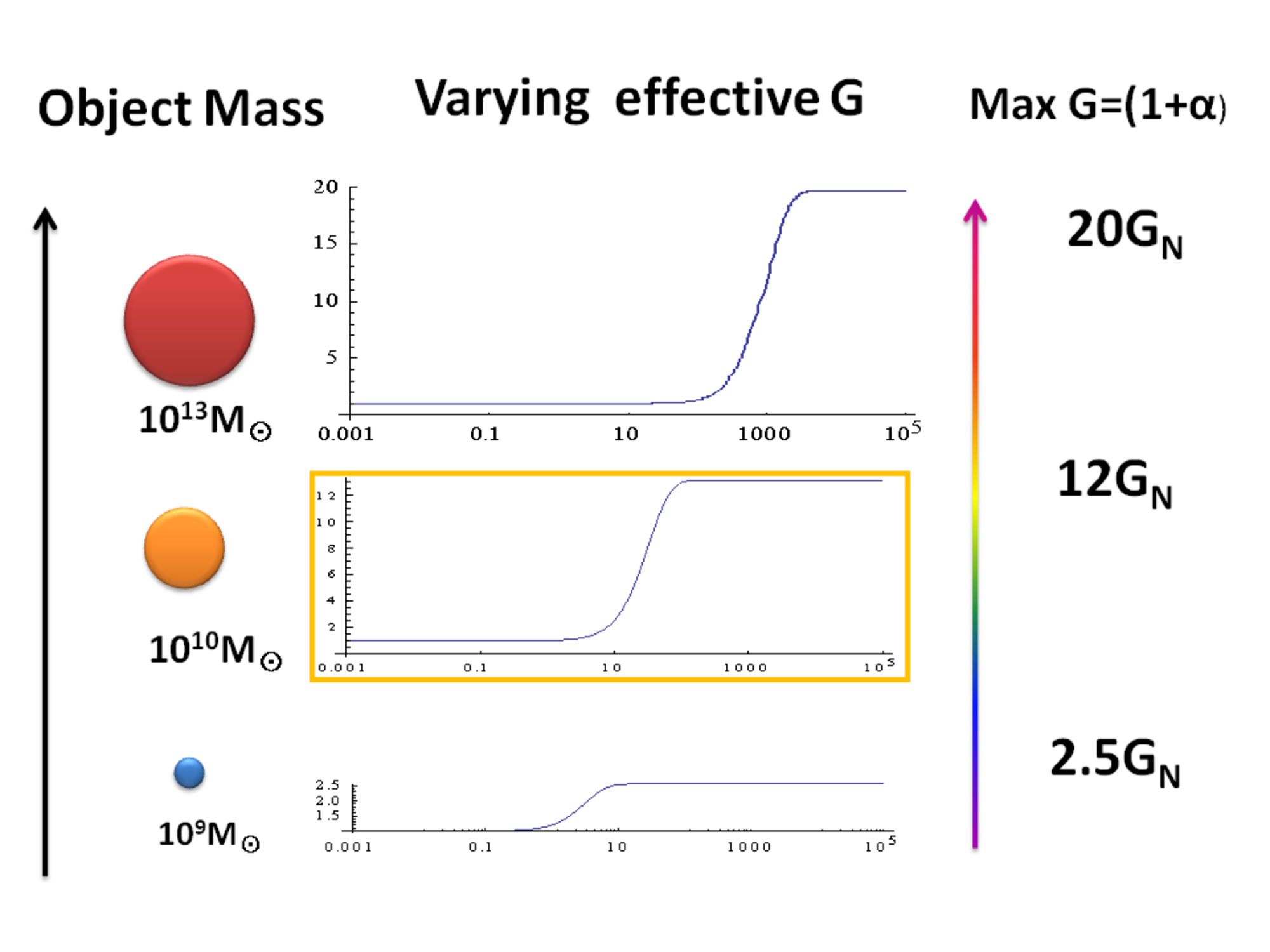}
\caption{Effective G and distance from the object.The example of three  different weight's objects:\newline$10^9 M_\odot$,$10^{10} M_\odot$,$10^{13} M_\odot$.}
\label{fig:P31_Geff_exp}
\end{figure}

MOG advocate that we recognize this strong gravity appearing by leaving for far away as dark matter.
$[1+\alpha-\alpha(1+r/\lambda)e^{-r/\lambda}]$ in Eq.\ref{PointAcceleration} is able to interpreted as effective gravitational constant.
Fig.\ref{fig:P31_ScaleVSG} shows relation between mass scale / distance scale and effective G.
The maximum of effective G and the effective range grow large depending on the mass of object.(See Fig. \ref{fig:P31_Geff_exp})
Therefore it can explain an effect of the dark matter of every scale.


\begin{figure}[h]
\centering
\includegraphics[keepaspectratio=true,width=13cm]{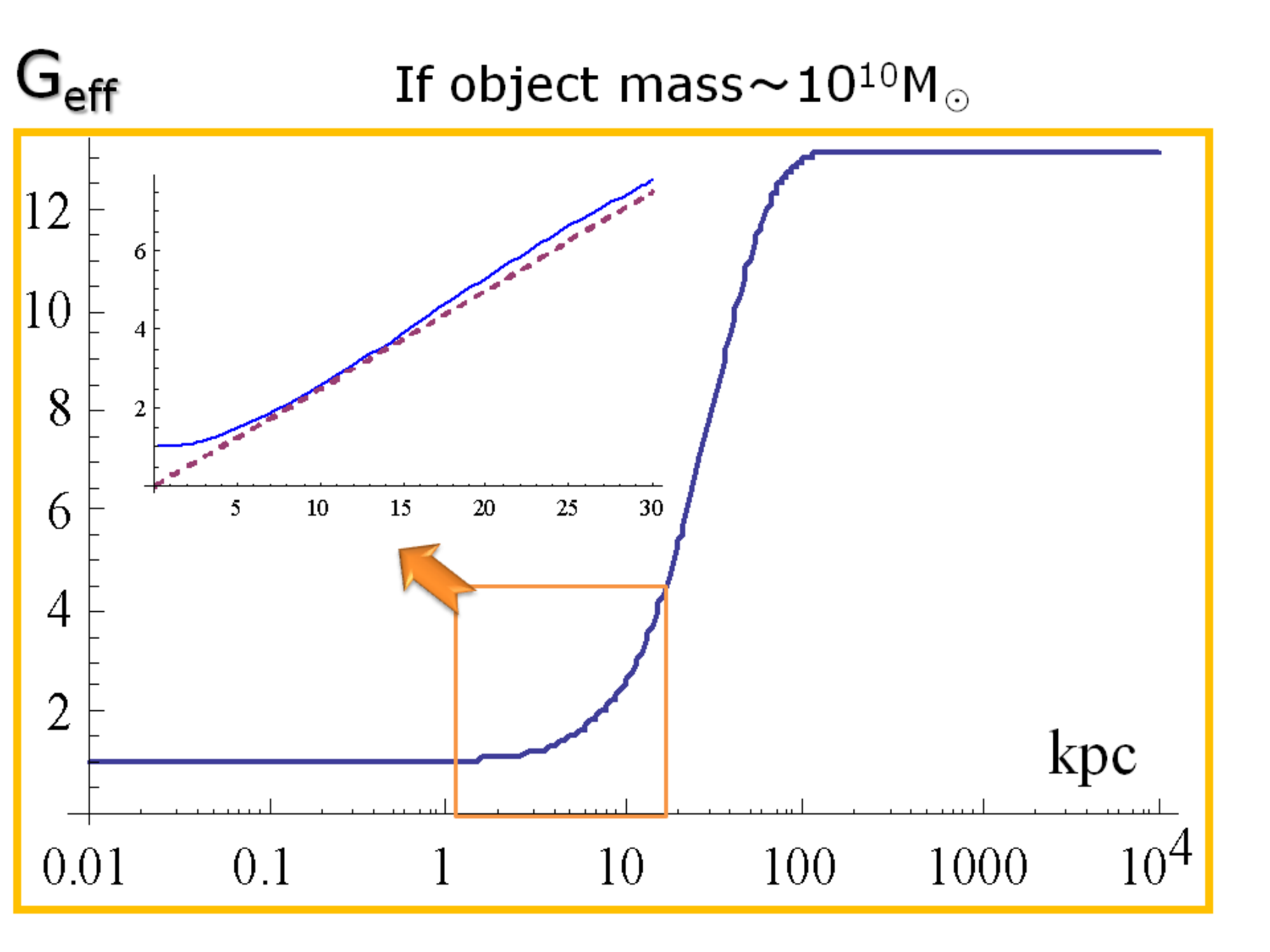}
\caption{Relation between effective G and distance from the object.\newline if object mass is $10^{10} M_\odot \sim$ galaxy disk,On disk scale,$G_{eff}$ is in proportion to r.}
\label{fig:P31_Geff10solarmass}
\end{figure}

The reason why the MOG can explain galactic flat rotary curves without dark matter halo is as follows.
Fig.\ref{fig:P31_Geff10solarmass} shows effective gravitatonal constant of $10^{10} M_\odot$(It is comparable to galaxy disk).
On galaxy disk scale ,$G_{eff}$ is in proportion to a radius.Let us remember equation of Kepler motion.
The revolution velocity of object having the circular orbitis is discribe as $v=\sqrt{\frac{GM}{r}}$.
If $G_{eff}$ is in proportion to r, velocity is constant.

The above is qualitative explanation, but agreement of observed rotation velocity by Doppler effect is identified as predicted rotation velocity by brightness in 100 galaxy\citep{Brownstein}\citep{Brownstein2}.

\section{Techniques and detail of N-body simulations}

The gravity acceleration equation of the STVG in a weak gravitational field is shown in the Eq.\ref{PointAcceleration}.
But,This equtaion is adaped only gravitational field of ``a point mass''.
It is not applied simple superposition principle of forces without self-contradiction.
The precedent study by Brandao \citep{Araujo} treat MOG as simple Yukawa gravity.Yukawa-effective range is fixed as $13.96 kpc$.
But,MOG is not simple Yukawa gravity.Yukawa-effective range(=$\lambda$)and increase of effective G(=$\alpha$) depends on neighboring mass distribution.
Every combination of an interacting particle(i-j) have $\lambda_{ji}$ and $\alpha_{ji}$.

The mutual gravity between the particle by STVG is written in Moffat et al.(2009)\cite{Toth5}.
I let an equation written in the paper become disintegration. 
And the following expressions are derived by adapting itself to many body system.
\begin{eqnarray}
a_j=-\sum_i^N \frac{G_N m_j (\vec{r_i}-\vec{r_j})}{(\left|r_i-r_j \right|^2+eps^2)^{3/2}}G_{eff}(i,j) \label{TwobodyAcceleration}
\end{eqnarray}
\begin{eqnarray}
G_{eff}(i,j)= \left[1+\alpha_{ij}-\alpha_{ij}(1+\frac{\left| r_i-r_j \right|}{\lambda_{ij}})e^{-\left| r_i-r_j \right|/\lambda_{ij}}\right]\label{Geff_ij}
\end{eqnarray}
\begin{eqnarray}
\lambda_{ij}=\frac{\sqrt{M_{eff}}}{D},\alpha_{ij}=\frac{19M_{eff}}{(\sqrt{M_{eff}}+E)^2},M_{eff}(i,j)=\sum_l^N m_l exp(-\xi \frac{\left| r_l-r_j \right|}{\left| r_i-r_j \right|})
\end{eqnarray}
where $\xi$ is a parameter which is decided by observation.
Even if the reader watches a numerical formula, the image of the physical meaning is hard to appear.
In the next page, a specific example is explained using a figure.

\newpage
\begin{figure}[h!]
\centering
\includegraphics[keepaspectratio=true,width=14cm]{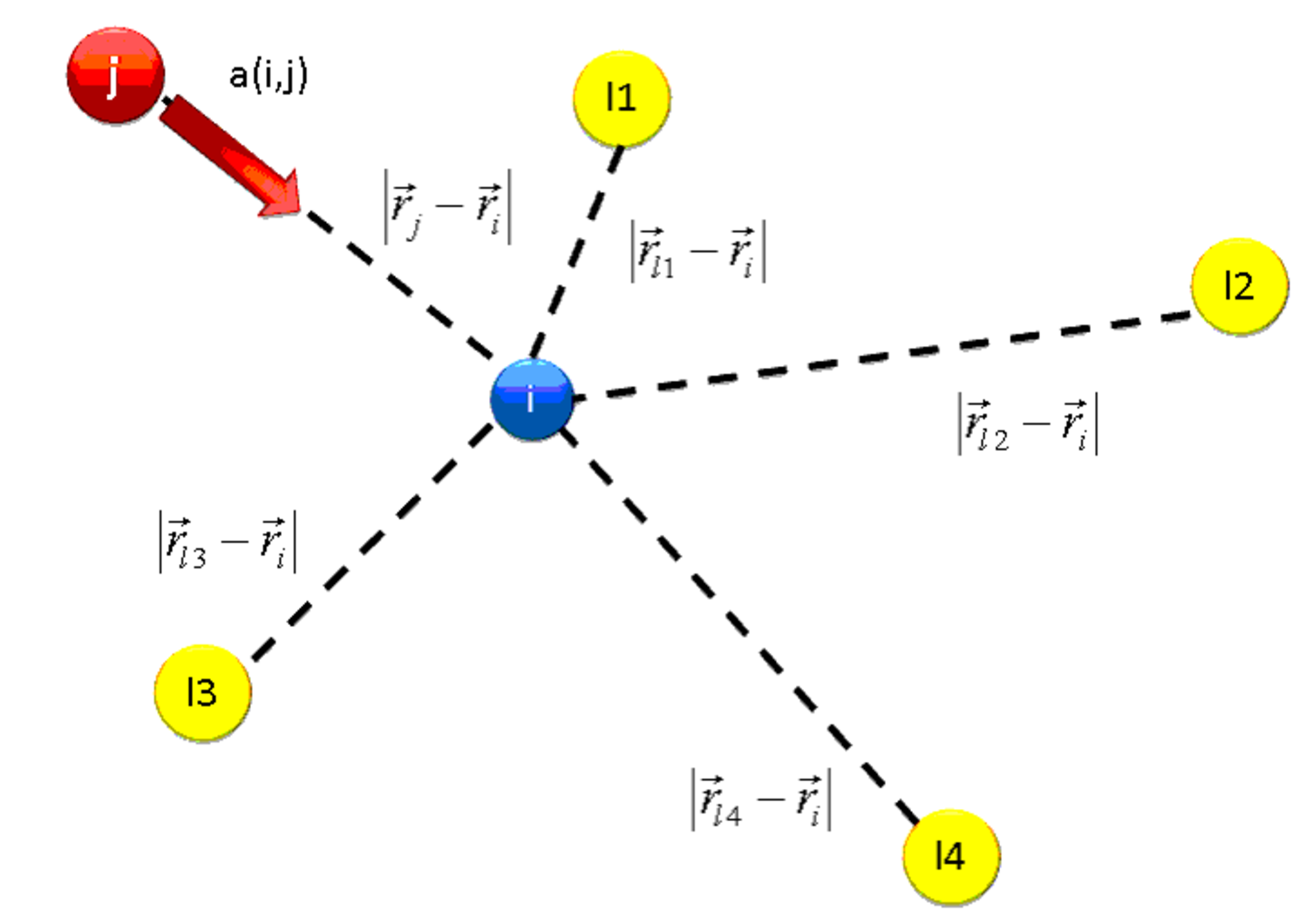}
\caption{Example of the acceleration calculation}
\label{fig:exact_example}
\end{figure}

We think about the placement of the particle such as the Fig.\ref{fig:exact_example}.
the gravitation that i particle gives to j particle is described in the following expressions: 

\begin{eqnarray}
a_{i,j}=- \frac{G_N m_j (\vec{r_i}-\vec{r_j})}{(\left|r_i-r_j \right|^2+eps^2)^{3/2}}G_{eff}(i,j) \label{TwobodyAcceleration ij}
\end{eqnarray}
\begin{eqnarray}
G_{eff}(i,j)= \left[1+\alpha_{ij}-\alpha_{ij}(1+\frac{\left| r_i-r_j \right|}{\lambda_{ji}})e^{-\left| r_i-r_j \right|/\lambda_{ji}}\right]\label{Geff_ij2}
\end{eqnarray}
\begin{eqnarray}
\lambda_{ij}=\frac{\sqrt{M_{eff}}}{D},\alpha_{ij}=\frac{19M_{eff}}{(\sqrt{M_{eff}}+E)^2}
\end{eqnarray}
\begin{eqnarray}
M_{eff}(i,j)=m_j+m_{l1} e^{(-\xi \frac{\left| r_{l1}-r_j \right|}{\left| r_j-r_i \right|})}+m_{l2} e^{(-\xi \frac{\left| r_{l2}-r_j \right|}{\left| r_j-r_i \right|})}+m_{l3} e^{(-\xi \frac{\left| r_{l3}-r_j \right|}{\left| r_j-r_i \right|})}+m_{l4} e^{(-\xi \frac{\left| r_{l4}-r_j \right|}{\left| r_j-r_i \right|})}
\end{eqnarray}

The calculation of gravitational acceleration on Moffat gravity needs the information of all the particles whenever we calculate force between two particles.
That is why the calculation number of times becomes $\mathcal{O}(N^3)$.By the normal method, we cannot carry out such an enormous calculations.

\newpage

\begin{figure}[h!]
\centering
\includegraphics[keepaspectratio=true,width=15cm]{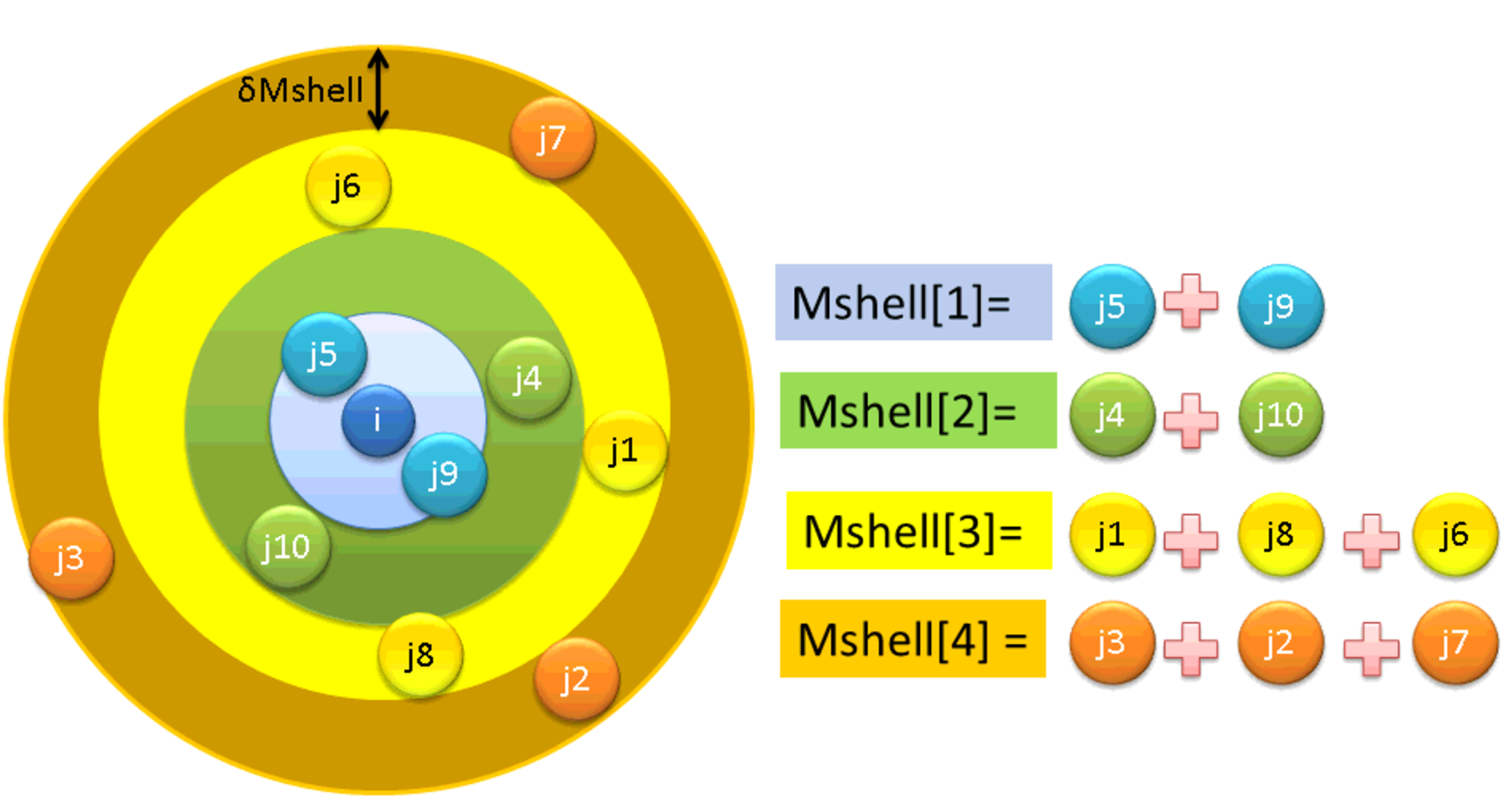}
\caption{Image of mass shell}
\label{fig:P31_Mshell}
\end{figure}

We show below the method of reducing calculation number of times.
We pay attention that $M_{eff}$ which decide the constant of gravitation increment($=\alpha$) and Yukawa-effective range($=\lambda$) depends on ``absolute value'' of the distance between the particle.
$M_{eff}$ integrated a function of the mass density per unit of area of the shell around the gravity source.
We can reduce calculation number of times by defining to an individual particle with the ``mass shell'' which is a state in  Fig.\ref{fig:P31_Mshell}.
In the following, we abbreviate ``mass shell'' to ``M shell''.

$G_{eff}$ don't depend on direction from the gravity source particle to the gravity receive particle ,but only distance from source particle to receive particle.
Therefore,we can define the $G_{eff}$'s shell(G shell) concentric from the gravity source particle.
We propose the following numerations from these hints.
In the case of normal N-body simulation,we calculate vector sum of net force on every gravity receive paritcle from all the other gravity source particles.
But,in the case of MOG,we calculate in the procedure that every gravity source particle gives force for all the other gravity receive particles.

\begin{enumerate}
 \item Getting mutual coordinates of all the other gravity receive particles from the gravity source particle.In parallel,defining``mass shell''.(The calculation number of times is $N-1$)
 \item Defining``G shell''by using ``mass shell''.It is necessary sum of all the ``mass shell'' to calculate one`` G shell''(The calculation number of times is $N mass shell \times N G shell$)
 \item The gravity source particle giving force for all the other gravity receive particles.Using ``G shell'' for this calculation.(The calculation number of times is $N-1$)
\end{enumerate}

We carry out this calculation for all the gravity source particles repeatedly.(The calculation number of times is $N$)
\newpage

Assuming that both ``mass shell''  and ``G shell''have equal interval shell,the basic equation is renewed below.
We define $\delta M shell$ as interval of ``mass shell'', and $\delta G shell$@as interval of ``G shell''.
\begin{eqnarray}
a_j=-\sum_i^N \frac{G_N m_j (\vec{r_j}-\vec{r_i})}{(\left|r_j-r_i \right|^2+eps^2)^{3/2}}G_{eff}(i,\lfloor \left|r_j-r_i \right|/\delta G shell \rfloor) \label{TwobodyAcceleration Gshell}
\end{eqnarray}
\begin{eqnarray}
G_{eff}shell(i,nGshell)= \left[1+\alpha(i,nGshell)-\alpha(i,nGshell)(1+\frac{\delta G shell \times nGshell}{\lambda(i,nGshell)})e^{-\frac{\delta G shell \times nGshell}{\lambda(i,nGshell)}}\right]\label{Gshell}
\end{eqnarray}
\begin{eqnarray}
\lambda(i,nGshell)=\frac{\sqrt{M_{eff}}}{D},\alpha(i,nGshell)=\frac{19M_{eff}}{(\sqrt{M_{eff}(i,nGshell)}+E)^2}
\end{eqnarray}
\begin{eqnarray}
M_{eff}(i,nGshell)=\sum_{n_{M shell}}^{N_{M shell}} M shell[n_{M shell}] exp(-\xi \frac{\delta M shell \times n_{M shell}}{\delta G shell \times nGshell})
\end{eqnarray}
\begin{eqnarray}
M shell(i,n_{M shell})=\sum_{j}^{N} \left\{
\begin{array}{ll}
m_j , &\quad (\lfloor \left|r_i-r_j \right|/\delta M shell \rfloor=  n_{M shell}) \\
0 , &\quad (otherwise)
\end{array}
\right.
\end{eqnarray}
where $\lfloor X \rfloor$ means rounding off fractions of $X$.For example,$\lfloor 1.2 \rfloor=1$,$\lfloor 3.6 \rfloor=3$.
In reality, ``mass shell''  and ``G shell''  does not necessarily have to be equal interval.Considering function form of $M_{eff}$,``mass shell'' of interval should be small in a center area.
Because the number of shell which we discribe in a program code is not infinity ,this method can adopt for only the system whose size does not change.
We treat a system of the dynamics equilibrium in most studies. Thus,we do not have any problem in this matter.

\begin{figure}[h!]
\centering
\includegraphics[keepaspectratio=true,width=14cm]{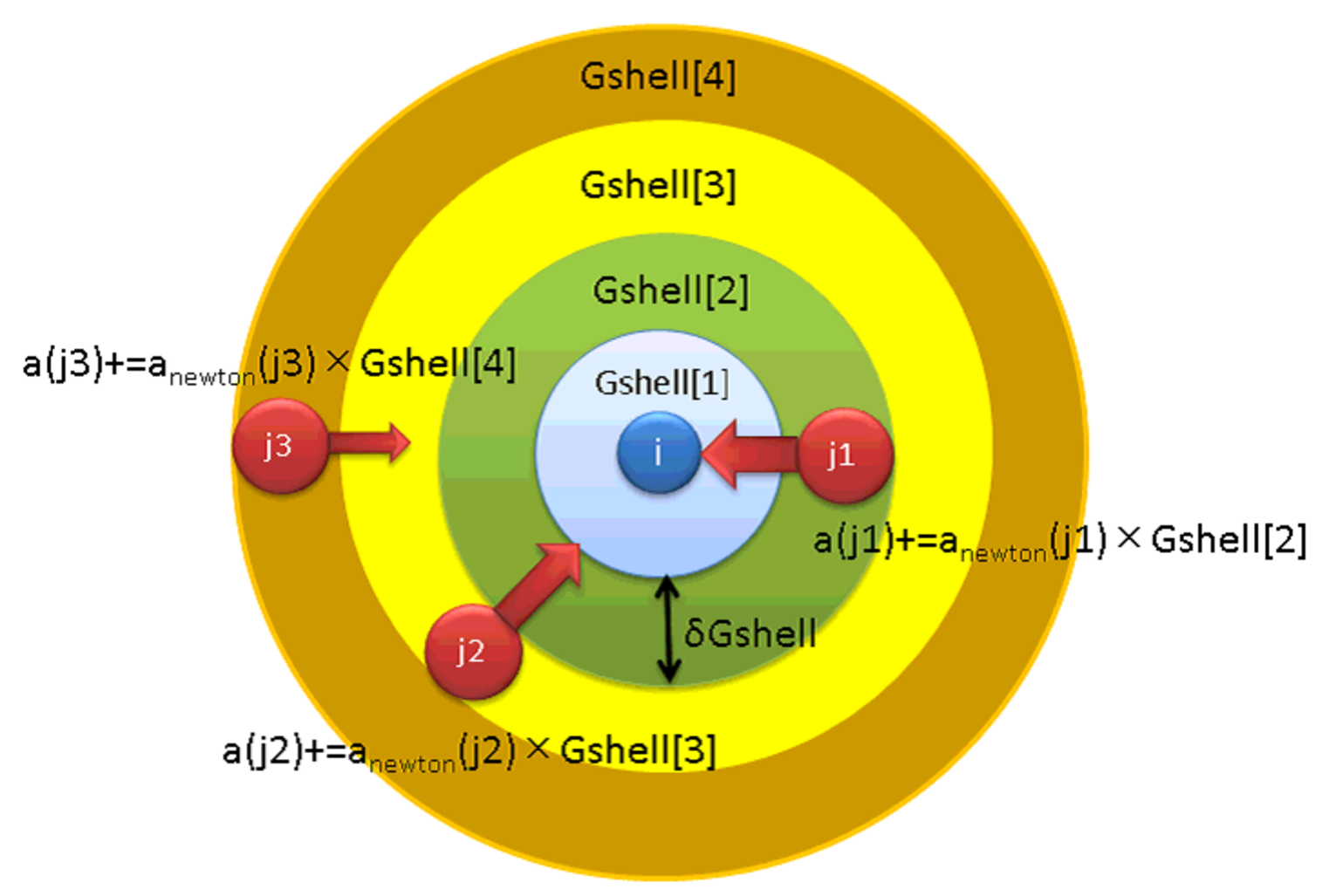}
\caption{Image of the acceleration calculation using ``G shell''}
\label{fig:Gshell_calc}
\end{figure}

\newpage

The precision of this approximate calculation depends on the number of ``mass shell'' and ``G shell''.
To examine dependence of the precision by number of ``shell'',we calculated the acceleration which each particle of the expnential disk received.The following is parameter of the disk.Thease are comparable to the Milky way.

\begin{table}[htbp]
\begin{center}
\begin{tabular}{ccc}
\hline
1R(radius) & 25kpc \\
1M(mass) & $5\times 10^{10} M_\odot$ \\
1T(time) & 0.26Gry \\
1V(velocity) & 92km/s \\
\hline
\end{tabular}
\caption{The standardization of the unit}
\end{center}
\end{table} 

\begin{table}[htbp]
\begin{center}
\begin{tabular}{ccc}
\hline
$R_d$(scale length of disk) & 2.5kpc \\
$R_{cut}$(radius of disk) & 25.0kpc \\
$Z_d$(scale length of disk height) & 0.5kpc \\
$M_d$(total mass of disk) & $5\times 10^{10} M_\odot$ \\
eps(softening length) & 0.048kpc \\
Nbody(number of particles) & 10000 \\
$\xi$(MOG parameter) & 12.0 \\
\hline
\end{tabular}
\caption{Setting parameters}
\end{center}
\end{table} 
The radius of most outer ``shell''(:$\delta M shell \times N_{M shell}=\delta G shell \times N_{G shell}$) is about $3.0$.
This length is enough to caluculate acceleration of the pair of particle which have the longest distance in the disk.
We examine the 7 model: $N_{M shell} \times N_{G shell}=10^2,50^2,100^2,250^2,500^2,1000^2,2000^2$.
For a comparison, we calculated the model of the $\mathcal{O}(N^3)$ exact calculation.

\begin{figure}[h!]
\centering
\includegraphics[keepaspectratio=true,width=14cm]{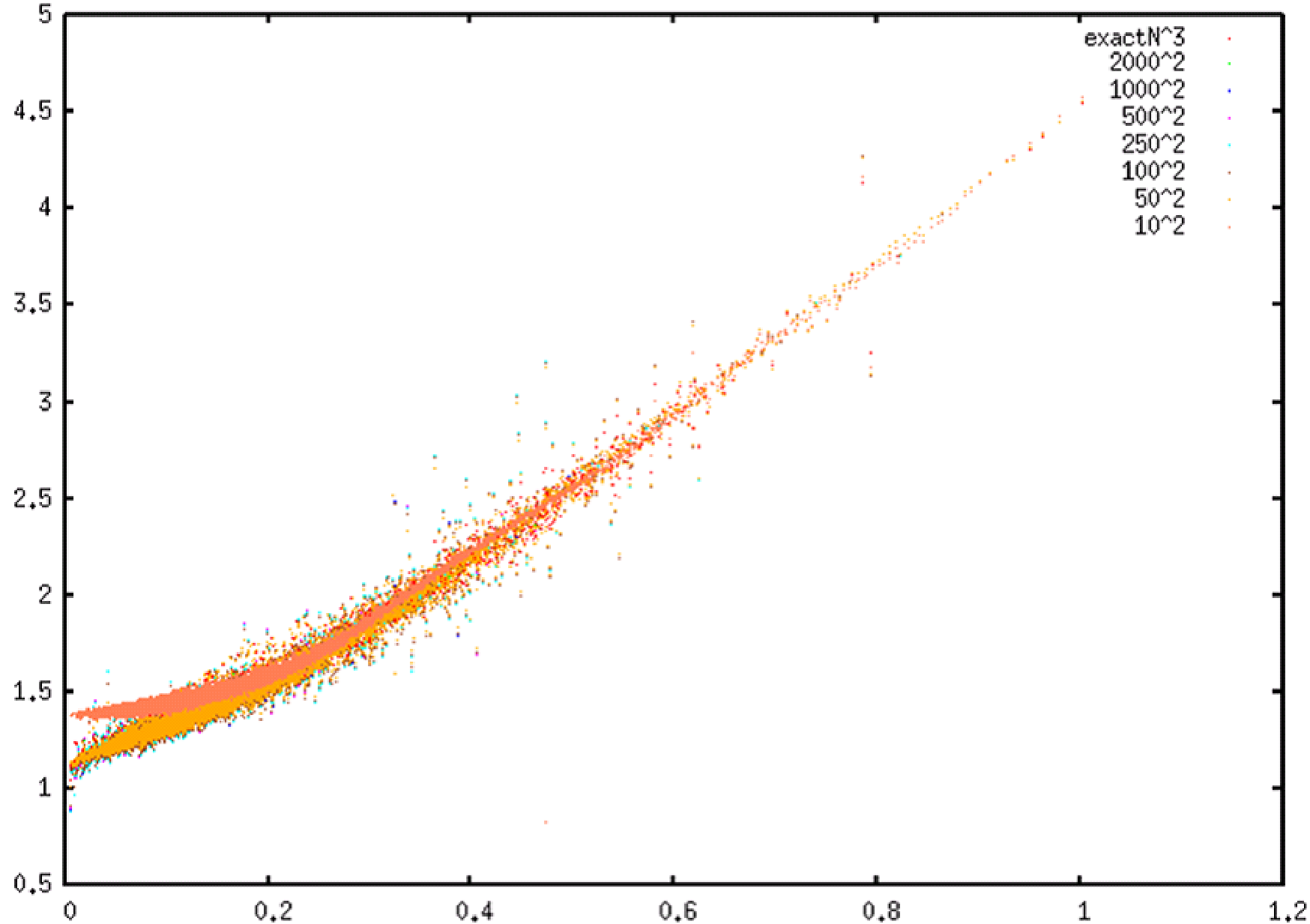}
\caption{The ratio of central acceleration by the modified gravity and the central acceleration by the Newton gravity.}
\label{fig:VS1}
\end{figure}

\begin{figure}[h!]
\centering
\includegraphics[keepaspectratio=true,width=14cm]{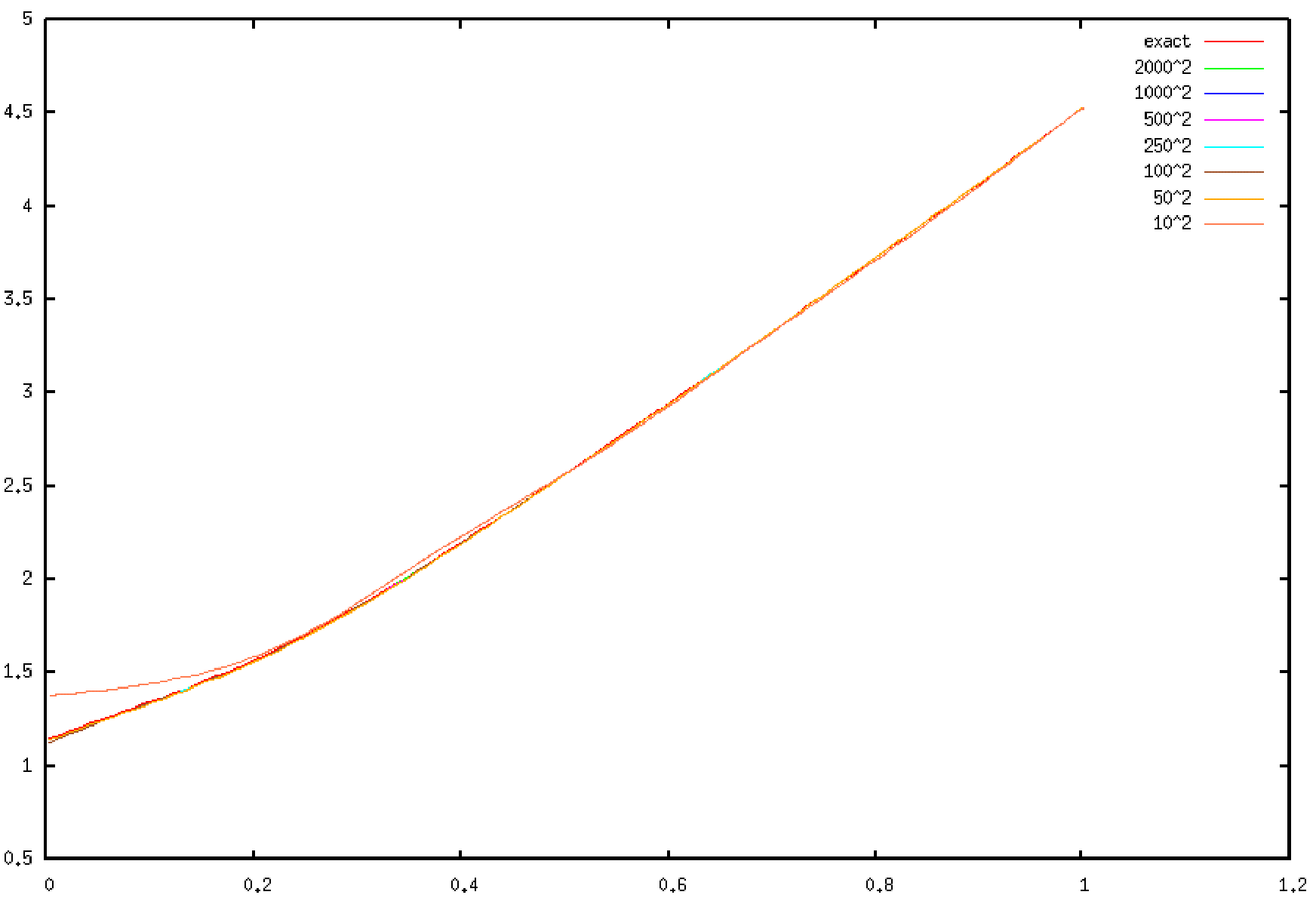}
\caption{lines which averaged each points in Fig 7.}
\label{fig:VS2}
\end{figure}

Then,we calculated the MOG acceleration of each particle in this disk.The result is shown in Fig.\ref{fig:VS1}.
This figure shows the ratio of central acceleration by the modified gravity and the central acceleration by the Newton gravity as an effective gravitational constant.
The data plots in the Fig\ref{fig:VS1} overlaps ,we cannot distinguish each model.
The Fig\ref{fig:VS2} expresses the lines which averaged each point in Fig.\ref{fig:VS1}.
Each line in the Fig\ref{fig:VS2} also overlaps ,we cannot distinguish each model.
However, it's the important that all other line except ``$10^2$ shell model'' overlaps with exact calculation model which has calculation number of times $\mathcal{O}(N^3)$.
After all, we don't need to divide interval of ``shell'' so much.
If  $N mass shell \times N G shell$ is smaller than $N$,the total calculation number of times becomes $\mathcal{O}(N^2)$.
In this case,the calculation load become equivalent to the normal N-body calculation.
We are groping for the method to reduce the calculation number of times more.

\section{Postscript}
This paper is an interim report.
Now,we are carrying out a large-scale calculation about galactic dynamics and large scale structure of universe.
Nurmerical calculation are carried out on Cray XT4 at Center for Computational Astrophysics(CfCA) of National Astronomical Observatory of Japan and SR16000 at YITP in Kyoto University.
If a meaningful result of the simulation appears, we intend to submit an official article.

\newpage

\newpage


\end{document}